\documentclass[shortnote,twocolumn]{jpsj2}
\usepackage{graphicx}
\def\runtitle{Isolated resonances in conductance fluctuations in ballistic billiards}
\def\runauthor{Achim \textsc{Manze}, Arnd \textsc{B\"acker}, Bodo \textsc{Huckestein} and Roland \textsc{Ketzmerick}}

\title{Isolated resonances in conductance fluctuations in ballistic billiards
}

\author{
Achim \textsc{Manze}$^1$\thanks{E-mail: amz@tp3.ruhr-uni-bochum.de}, Arnd \textsc{B\"acker}$^2$, Bodo \textsc{Huckestein}$^1$ Roland \textsc{Ketzmerick}$^3$
}

\inst{%
$^1$ Institut f\"ur Theoretische Physik III, Ruhr-Universit\"at
  Bochum, D-44780 Bochum, Germany\\
$^2$ Abteilung Theoretische Physik, 
  Universit\"at Ulm, Albert-Einstein-Allee 11, D-89081 Ulm, Germany\\
$^3$ Max-Planck-Institut f\"ur Str\"omungsforschung and Institut
  f\"ur Nichtlineare Dynamik der Universit\"at G\"ottingen, Bunsenstra{\ss}e 10, D-37073 G\"ottingen, Germany
}

\recdate{\today}

\abst
{
}

\kword{%
soft quantum chaos, semiclassical theories, scattering, billiard system 
}

\begin{document}
\sloppy
\maketitle

\section{Introduction}
One possibility to study quantum mechanical manifestations of classical chaos is the calculation of transport properties. A well known example is the
occurrence of the universal conductance fluctuations in billiards
whose classical counterparts show completely chaotic
dynamics~\cite{Jal2000}. However, this is the extreme opposed to integrable dynamics. Generic systems have a mixed phase space, where regions of regular
motion coexist with chaotic ones~\cite{LicLie92}. For the quantum mechanical analog of such systems it was derived semiclassically that the graph $G$ vs control parameter should be a fractal \cite{Ket96}. 

Surprisingly, for the cosine billiard, a system that can exhibit a mixed phase space, a recent numerical study did not show these fractal conductance fluctuations but sharp isolated resonances of the conductance and the Wigner-Smith time delay with a width distribution covering several orders of magnitude~\cite{HucKetLew99}. 

Here we focus on the resonances of the Wigner-Smith time which are of Breit-Wigner shape.  We can clarify the situation by an analysis of scattering states of the open cosine billiard as well as eigenstates of the closed billiard. A phase space representation of these is used to classify them as chaotic, hierarchical or regular~\cite{KetHufSteWei2000}.
 We find evidence for the conjecture  that the isolated resonances are related to the hierarchical and regular eigenstates of the closed system \cite{HufWeiKet2001}. 
\section{Model and strategy}

\begin{figure}[b]
  \begin{center}\leavevmode
    \includegraphics[width=0.8\linewidth]{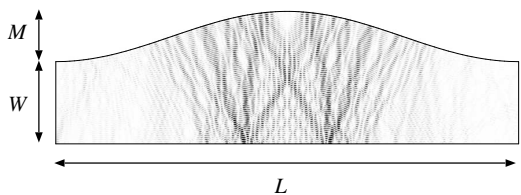}
    \caption{The cosine billiard for the parameters $W/L=0.18$ and
      $M/L=0.11$. Inside the scattering state of a hierarchical
      state is shown.}
    \label{fig:1}
  \end{center}
\end{figure}
We study the cosine billiard~\cite{HucKetLew99} , either closed (boundary at $x=0$ and $x=L$) or with semi-infinite leads attached to both sides. The boundary consists of the line $y=0$ and
\begin{equation}
  \label{eq:2}
  y(x)=W+\frac{M}{2}\left[1-\cos\left(\frac{2\pi x}{L}\right)\right]
\end{equation}
for $0\le x\le L$ (see Fig. 1). The classical phase space structure can be tuned by varying the ratios $W/L$ and $M/L$. For $W/L=0.18$ and $M/L=0.11$ it is a mixed phase space.  

\begin{figure}[tb]
  \begin{center}\leavevmode
    \includegraphics[width=0.9\linewidth]{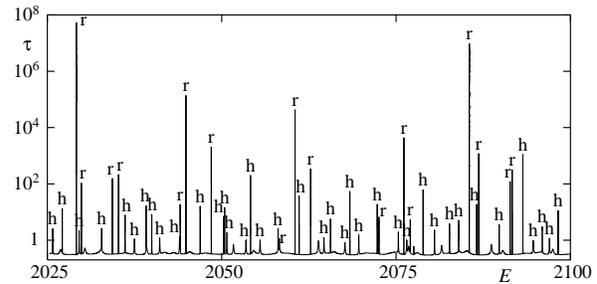}
    \caption{Wigner-Smith delay time $\tau$ vs energy $E$ (energy unit:
      $E_0=\hbar^2 \pi^2 / (2 m W^2)$). The labels of the resonances
      indicate whether they correspond to regular (r) or hierarchical
      (h) eigenstates of the closed system.}
    \label{fig:2}
  \end{center}
\end{figure}

In this case sharp resonances in the Wigner-Smith time delay
$\tau=\frac{-i\hbar}{2N} \mbox{Tr}\,(S^\dagger dS/dE)$ of the system
($2N$ is the dimension of the $S$-matrix) occur as can be seen in
Fig.~\ref{fig:2}. The calculation of $S$ is outlined
elsewhere~\cite{HucKetLew99,bhk01}.

The Husimi representation $H_n(s,p)$ of the eigenstates is used to estimate the coupling of the eigenstates to the leads by integrating $H(s,p)$ along the opening
\begin{equation}
  \label{eq:eta}
  \eta_n=\int_{-W}^0\textrm{d}s\int_{-1}^{1}\textrm{d}p\,H_n(s,p)\,.
\end{equation}
This quantity was defined in the context of the definition of hierarchical states~\cite{KetHufSteWei2000}. It should be proportional to the width $\Gamma$ of a resonance. 

\begin{figure}[t]
  \begin{center}
    \leavevmode
    \includegraphics[width=0.8\linewidth]{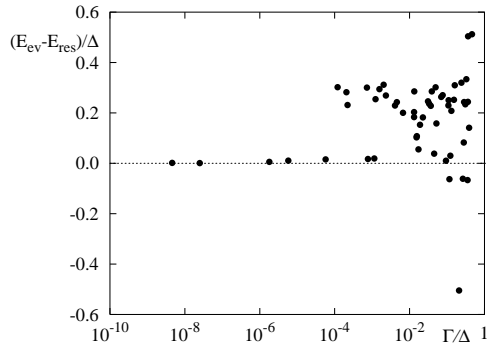}

    \caption{Difference of eigenstate energy $E_{\rm ev}$
        and resonance energy $E_{\rm res}$ in units of the mean
        level spacing $\Delta$ vs $\Gamma/\Delta$.
        The deviations increase with $\Gamma$.}
    \label{fig:DeltaE-vs-Gamma}
  \end{center}
\end{figure}
\begin{figure}[t]
  \begin{center}
    \leavevmode
    \includegraphics[width=0.8\linewidth]{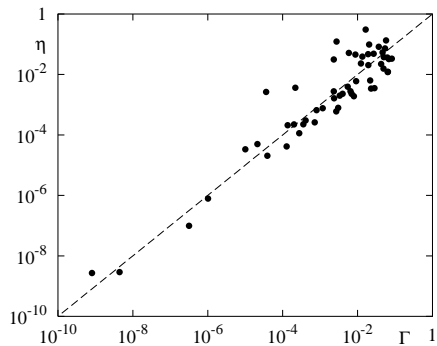}

    \caption{The strength $\eta$ of an eigenstate at the
        left boundary vs the resonance width $\Gamma$
        of the corresponding resonance.
        An approximate proportionality can be seen.}
    \label{fig:eta-vs-Gamma}
  \end{center}
\end{figure}

\section{Results}
\begin{figure}[bt]
  \begin{center}
    \leavevmode
    \includegraphics[width=0.8\linewidth]{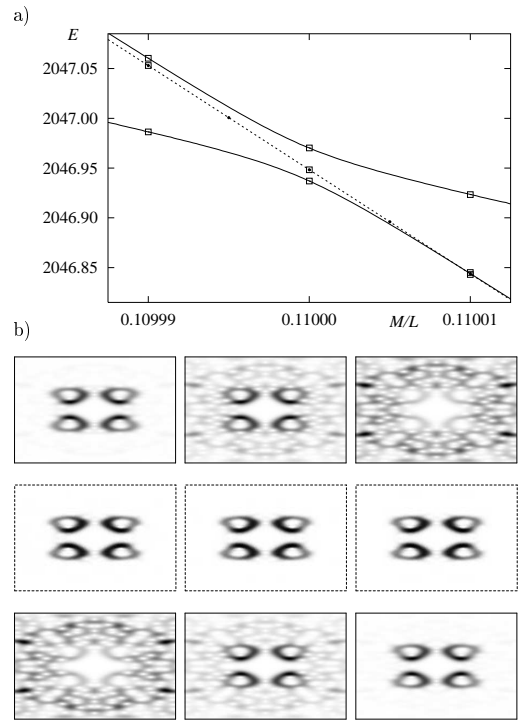}
    \caption{(a) Energies of states $5736$ and $5737$ of odd symmetry (solid
        lines) showing an avoided crossing under variation of $M/L$.
        The energy of the only isolated resonance in this energy range
        (dots connected by a dashed line) follows the regular state of
        the closed system.  (b) The Husimi representations for odd state $5737$ 
        (top row), the
        scattering state (middle row) and for odd eigenstate $5736$
        (bottom row) are shown for $M/L=0.10999,\,0.11,\,0.11001$ (left
        to right). For the
        eigenstates one clearly sees the typical exchange of the
        structure while passing the avoided crossing whereas the
       scattering state is not affected.}
    \label{fig:paramvar}
  \end{center}
\end{figure}

It is reasonable to assume that eigenstates leading to a given resonance at energy $E_{\textrm{res}}$ have energy $E_n$ close to $E_{\textrm{res}}$. A  posteriori this is illustrated in Fig.~\ref{fig:DeltaE-vs-Gamma} where one can see that most of the resonances are not farther away from the corresponding eigenenergy than about 30~\% of the mean level spacing $\Delta$. As an additional assumption we take $\eta_n\propto\Gamma_{\textrm{res}}$  and are able to associate most of the resonances to eigenenergies of the closed system, which is illustrated in Fig.~\ref{fig:eta-vs-Gamma}. The same assignment can be made by visually comparing the Husimi representation of eigenstates of the closed system with a corresponding phase space representation of the scattering states~\cite{bhk01}. This can assist the former method as there are deviations from the linear relationship between $\eta_n$ and $\Gamma_{\textrm{res}}$. Comparatively large deviations may occur due to avoided crossings of the closed system (see Fig.~\ref{fig:paramvar}). In the  region of an avoided crossing of a e.\,g. regular and a chaotic eigenstate the Husimi representation shows regular as well as chaotic contributions. Hence the quantity $\eta_n$ need not be of the same order of magnitude as $\Gamma_{\textrm{res}}$ in these cases.    

\section{Conclusion}
We have shown that sharp resonances in the transport properties of quantum systems with a mixed phase space are due to states concentrated in the regular and hierarchical parts of phase space. We found that scattering states can be associated to eigenstates via their phase space portraits which is important because open systems are experimentally more accessible whilst closed systems are preferred for numerical simulations.

\section*{Acknowledgements}
A.B.\ acknowledges support by the 
Deutsche Forschungs\-ge\-mein\-schaft under contract No. DFG-Ba 1973/1-1.

\end{document}